\documentclass[aps,prb,twocolumn,superscriptaddress,floatfix]{revtex4}

\usepackage{epsfig,amsmath,amsfonts,amssymb}

\def\carbon{$^{13}C$}

\def\sasf{$(TMTTF)_2AsF_6$}
\def\ssbf{$(TMTTF)_2SbF_6$}
\def\CiT1{$^{13}T_1^{-1}$}
\def\PiT1{$^{1}T_1^{-1}$}
\def\a{${\mathbf{a}}$}
\def\F19{$^{19}F$}

\def\prl{Phys. Rev. Lett.}
\def\prb{Phys. Rev. B}
\def\jpsj{J Phys. Soc. Japan}

\begin{document}

\title{Electron-lattice coupling and the broken symmetries of the molecular salt (TMTTF)$_2$SbF$_6$}

\author{W. Yu}
\email[]{weiqiang@physics.ucla.edu}
\affiliation{Department of Physics and Astronomy, University of California, Los Angeles, CA 90095}

\author{F. Zhang}
\affiliation{Department of Physics and Astronomy, University of California, Los Angeles, CA 90095}

\author{F. Zamborszky}
\affiliation{Department of Physics and Astronomy, University of California, Los Angeles, CA 90095}

\author{B. Alavi}
\affiliation{Department of Physics and Astronomy, University of California, Los Angeles, CA 90095}

\author{A. Baur}
\affiliation{Department of Chemistry and Biochemistry, University of California, Los Angeles, CA 90095}

\author{C. A. Merlic}
\affiliation{Department of Chemistry and Biochemistry, University of California, Los Angeles, CA 90095}

\author{S. E. Brown}
\affiliation{Department of Physics and Astronomy, University of California, Los Angeles, CA 90095}

\date{\today}

\begin{abstract}
(TMTTF)$_2$SbF$_6$ is known to undergo a charge ordering (CO) phase transition at $T_{CO}\approx156K$ and another 
transition to an antiferromagnetic (AF) state at $T_N\approx 8K$. Applied pressure $P$ causes a decrease in both $T_{CO}$ 
and $T_N$. When $P>0.5 GPa$, the CO is largely supressed, and there is no remaining signature of AF order. Instead, the 
ground state is a singlet. In addition to establishing an expanded, general phase diagram for the physics of TMTTF salts, 
we establish the role of electron-lattice coupling in determining how the system evolves with pressure. 
\end{abstract}


\maketitle

The isostructural family of charge transfer salts $(TMTTF)_2X$ and $(TMTSF)_2X$ are formed with singly charged anions, 
such as $ClO_4^-$, $PF_6^-$, and $Br^-$ \cite{Jerome2002}, so the average hole count is $0.5$/donor. In the case of the 
$TMTTF$ salts, they are susceptible to a charge-ordering (CO) transition at temperatures of the order of $100K$ 
\cite{Chow2000,Monceau2001}, which is often attributed to the importance of both on-site and near-neighbor Coulomb 
repulsion \cite{Seo1997}, and influenced by electron-lattice coupling 
\cite{Mazumdar1999,Riera2001,Monceau2001,Clay2003,Brazovskii2003}. Compared to the analog $TMTSF$ materials, the 
bandwidths are much smaller, and therefore models naturally producing charge order and including only electronic degrees 
of freedom could be expected to describe some aspects of the physics correctly. Nevertheless, it is unclear whether 
realistic parameters successfully describe the experiments in several ways. First, it has been argued that the 
near-neighbor repulsion may not be strong enough to stabilize the charge order \cite{Mazumdar2000,Riera2001,Clay2003}. 
Furthermore, although the charge-order (CO) order parameter has not been determined directly, there is indirect evidence 
from transport \cite{Nad2002} and EPR \cite{Nakamura2003} measurements that the order-parameter's wavevector changes when 
the symmetry of the counterion is changed. An explanation should involve coupling of the charge degrees of freedom on the 
molecular stacks to the lattice. Calculations on one dimensional models including intramolecular, intermolecular, and 
counterion coupling indicate that when these degrees of freedom are included, a variety of new broken symmetry states are 
possible \cite{Clay2003,Riera2001,Brazovskii2003}. At least in the case of the insulating TMTTF materials, it is not be 
surprising that the robustness of the CO influences what ground state is observed \cite{Zamborszky2002}. To date, very 
little is known about the details of the observed phases and what controls their stability. 

\begin{figure}[htb]
\includegraphics[width=3.2in]{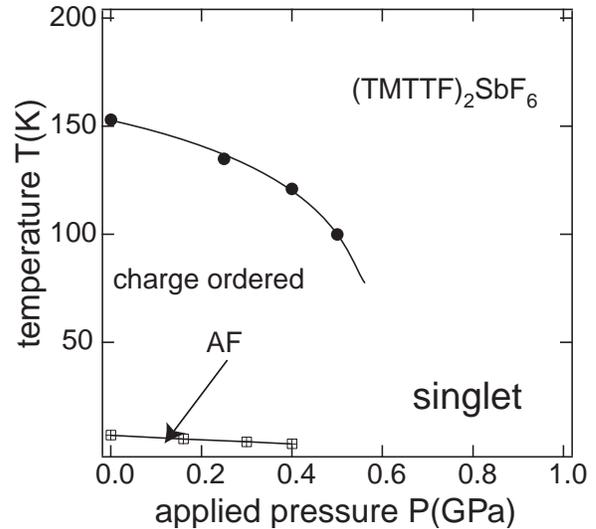}
\caption{ Pressure vs. temperature phase diagram for $(TMTTF)_2SbF_6$, identified using $^{13}C$ NMR spectroscopy. }
\label{SsbfPD}
\end{figure}

The sensitivity to chemical or mechanical pressure of this class of materials provides an opportunity to explore some 
general trends. The pressure/temperature phase diagram for \ssbf\ \cite{Yu2004}, determined using \carbon\ NMR 
spectroscopy on spin-labeled crystals, appears in Fig. \ref{SsbfPD}. As the pressure is increased, the ordering 
temperature $T_{CO}$ to the CO state decreases. $T_{CO}$ is reduced by almost half, to $90K$, with $0.5GPa$ applied 
pressure. Over the same pressure range, the charge order amplitude is significantly reduced, and becomes unidentifiable at 
pressures beyond it. As a consequence, the actual transition line is not established beyond $0.5GPa$. Also decreasing is 
the AF ordering temperature $T_N$. When $P>0.4 GPa$, no experimental signature for the AF state is observed. For technical 
reasons associated with the experiments that are likely complicated by quenched disorder, we have not identified a phase 
boundary for the ground state found at high pressures. Rather, the results of experiments conducted in the range of 
$T=2-5K$ exhibit the signatures of a singlet ground state \cite{Yu2004}. Below, we describe experiments that demonstrate 
the role of the counterion sublattice in determining the observed phase diagram.

The samples were prepared as described previously. Spin-labeled TMTTF donors were synthesized with two \carbon\ nuclei on 
the bridging sites at the center of the dimer molecules \cite{Merlic1999}, and subsequent crystal growth was carried out 
by electrolysis. The experiments consist of \carbon\ and \F19\ NMR spectroscopy on \ssbf\ as a function of pressure and 
temperature. In the first case, the external field was $B_0=9.0T$, with the field applied perpendicular to the \a\ 
(molecular stacking) axis. For the \F19\ measurements, $B_0=4.9T$ was used. High pressure experiments were performed using 
a standard BeCu clamp cell using Flourinert 75 (3M) for the medium. The stated pressure, as shown in Fig. \ref{SsbfPD}, is 
derived from the forces applied at $T=300K$, and systematic consistency was verified using separate calibration runs. 

\begin{figure}[htb]
\includegraphics[width=3.2in]{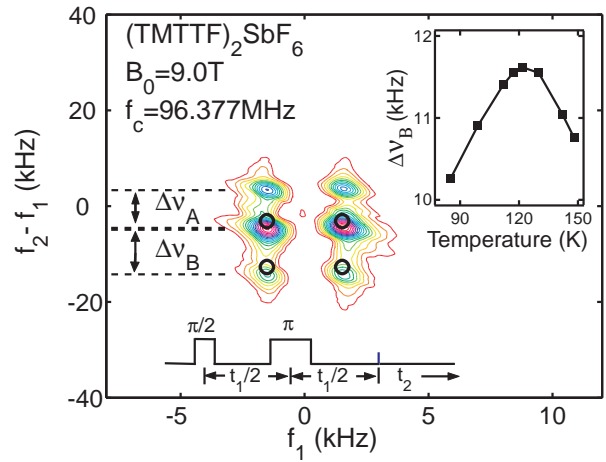}
\caption{\label{SSbFchargeOrder} \carbon\ NMR spectra at temperatures $T=85K<T_{CO}$ (contours) and $T=156K>T_{CO}$ 
(signal location represented by dark, open circles). The applied field is $9.0T$ directed in the $\mathbf{b'}-\mathbf{c*}$ 
plane. In the inset is the temperature dependence of the frequency difference ($\Delta\nu$) of the peaks.}
\end{figure}

Typical spectral changes are illustrated in Fig. \ref{SSbFchargeOrder}, which shows the results of two-dimensional 
spin-echo spectroscopy \cite{Ernst1990} \carbon\ NMR experiments. The experiment separates the hyperfine and chemical 
shifts from the internuclear dipolar coupling frequencies between pairs of neighboring \carbon\ nuclei 
\cite{approximation}. The shifts are plotted on the vertical $f_2-f_1$ axis, and the \carbon-dipolar coupling is shown on 
the horizontal $f_1$ axis. We note that the frequencies $f_1$ and $f_2$ result from the Fourier transform of the two 
dimensional data set in $(t_1,t_2)$, as defined in the figure.  The spectrum taken with $T<T_{CO}$ is shown directly, and 
the spectrum for $T>T_{CO}$ is represented by the dark, open circles. Consider first the spectrum of the high-temperature 
phase. The signal appears at two frequencies on the $f_2-f_1$ axis. When $T<T_{CO}$, the signal features multiply. 

We understand the spectrum and how the changes pertain to charge disproportionation as follows \cite{Chow2000}. Within 
each molecule, there are two inequivalent \carbon\ sites, with distinct hyperfine couplings. This is the reason for 
signals appearing at two frequencies along the $f_2-f_1$ axis in the high symmetry phase. Let's refer to them as $\nu_A 
(\nu_B)$ for the higher (lower) frequency. The number of features are {\it doubled} in the CO phase because two 
inequivalent molecular environments develop, which we can label as $\alpha$, $\beta$. As the NMR frequencies are related 
to the carrier densities on the $\alpha$ and $\beta$ molecules, the frequency differences ({\it 
e.g.},$\Delta\nu_A=\nu_{A\alpha}-\nu_{A\beta}$) are related to the CO order parameter. Some of the absorption features are 
unresolved in the spectrum shown. The inset shows the temperature dependence $\Delta\nu_B$, which saturates and then 
decreases upon cooling.

\begin{figure}[htb]
\includegraphics[width=3.2in]{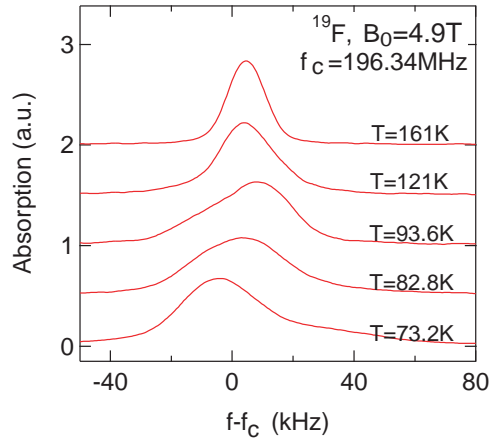}
\caption{Ambient pressure $^{19}F$ NMR spectra, recorded at different
temperatures.}
\label{F19spectra}
\end{figure}

\begin{figure}[htb]
\includegraphics[width=3.2in]{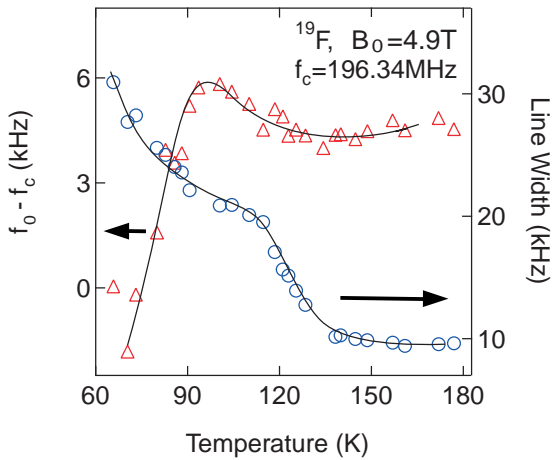}
\caption{First moment ($^{19}nu_0$) and linewidth, evaluated from $^{19}F$ spectra as shown in Fig. \ref{F19spectra}. The 
first moment is measured relative to $f_0$; see the text for a definition of the linewidth. The solid lines serve as 
guides to the eye.}
\label{F19moments}
\end{figure}

In Fig. \ref{F19spectra} are shown \F19\ spectra recorded over a range of temperatures. From approximately $T=130K$ and 
below, changes in both the first moment and linewidth are noticeable. These are summarized in Fig. \ref{F19moments}, where 
the linewidth is defined as the frequency span that includes half of the integrated intensity. Changes appear to occur in 
two steps, the first in a temperature range centered about $T=120K$ and the second in a temperature range around 
$T=75-80K$. 

In discussing these results, we first consider the nature of the \F19\ linewidth broadening. It is natural, at least in 
part, to consider the spectral effects as related to motion of the $SbF_6^-$ counterion. More generally, the counterions 
of the TMTTF and TMTSF salts fall into two classes: centrosymmetric and non-centrosymmetric. In the first class are the 
hexaflourides $PF_6^-$, $AsF_6^-$, and $SbF_6^-$, and $Br^-$, and in the second are $ClO_4$, $FSO_3$, etc. All are small 
and compact; $SbF_6^-$ is one of the largest, with the $F-Sb-F$ linear distance approximately $3.8\AA$. At high 
temperatures the orientation of the counterions is known to be highly disordered, and thought to be rotating 
\cite{McBrierty1982}. Upon cooling, the non-centrosymmetric counterions orientationally order \cite{Pouget1996} in a first 
order phase transition, lowering the space group symmetry of the crystal. The centrosymmetric ions are not reported to do 
that; instead, their motions are considered activated so linebroadening is expected to occur as a crossover upon cooling 
and there is no broken symmetry. At first glance, our results appear inconsistent with this scenario because the lineshape 
is asymetric. In a single crystal, this would indicate highly disordered sites at low temperatures, and the resulting 
variation of chemical or Knight shifts. From the data, we see that a distribution of chemical shifts is not observable 
(the line is homogeneously broadened). Nevertheless, we do not rule out crystal twins as a contributing factor in 
producing the observed lineshape. In either case, the broadening results when anionic motions become suitably slow or a 
first-order orientational ordering transition of some kind takes place. The second step in the \F19\ linewidth appears to 
be associated with dynamics of some of the methyl groups and $^1H-^{19}F$ coupling. As evidence, we note that there is a 
strong peak in the $^1H$ spin lattice relaxation in the same temperature range, and very similar in strength and 
temperature range to what is observed in the TMTSF salts \cite{McBrierty1982}. 

In this context, there is a natural association of the decrease in CO order parameter, from Fig. \ref{SSbFchargeOrder}, 
with the \F19\ line broadening and shift. For the broadening to occur, the anions must be stationary on a time scale of 
the order of the inverse (high-temperature) homogeneous linewidth \cite{Slichter1978}, which can occur through an 
activated diffusion process or as a result of a structural phase transition. For either case, the behavior of the order 
parameter suggests that suppression of counterion motion, or disorder associated with it, leads to suppression of the 
charge disproportionation on the donor molecules. 

Finally, we address the evolution with pressure of both the CO order parameter and the ground states. In Fig. 
\ref{SsbfPD}, neither the line demarking a transition to the CO phase, nor the CO/AF line is followed to $T=0$. In 
transport experiments \cite{Monceau2003}, there is an indication that the dielectric and resistive anomalies associated 
with the transition to the CO phase are monotonically suppressed with applied pressure. Our own transport studies confirm 
this. In the \carbon\ NMR spectrum, we observe some broadening of the spectral features, along with a weakening of the CO 
order parameter as pressure is increased. Beyond $P\approx0.5 GPa$, the CO features are unresolved. At the same time, the 
temperature at which $^{121}Sb$ linebroadening is observed coincides with the \F19\ broadening and {\it increases} with 
pressure \cite{Yu2004b}. These observations suggest that there is a phase competing with the CO, and its stability is 
associated with counterion degrees of freedom. 

 A natural first step is the quasi-one dimensional extended Hubbard model, including only electronic degrees of freedom  
\cite{Seo1997}. With large enough on-site and near-neighbor repulsions $U$ and $V$ (relative to hopping integral $t$) in 
$1/4$-filled systems, a charge pattern of alternating rich and poor sites is produced and the ground state is 
antiferromagnetic. Nevertheless, there are no diffraction experiments identifying the order parameter, and including 
electron-lattice coupling leads to other possibilities for the charge configuration of the CO state \cite{Mazumdar2000}. 
It is also argued that the physically appropriate values for $V$ may not exceed the threshold for charge order. 
Nevertheless, producing ferroelectricity, as inferred from low-frequency dielectric experiments \cite{Monceau2001}, from 
the CO phase requires more couplings. Including the counterion naturally gives rise to ferroelectric order parameters 
\cite{Riera2001,Brazovskii2003}. And Peierls-type coupling to the lattice leads to order parameters reminiscent of 
Spin-Peierls order \cite{Mazumdar1999,Riera2000}. Then, it follows that a large-amplitude CO follows from coupling to the 
counterion and competes with the Peierls coupling and the SP ground state \cite{Kuwabara2003}. Experimental evidence for 
the competition was seen previously in experiments on \sasf\ \cite{Zamborszky2002}. In the experiments reported here, the 
relative importance of the anion coupling is diminished once the anions freeze into place. That is, the freezing limits in 
some way the motion of the counterion. In turn, the coupling is reduced and the CO phase is destabilized.

\begin{figure}[htb]
\includegraphics[width=3.2in]{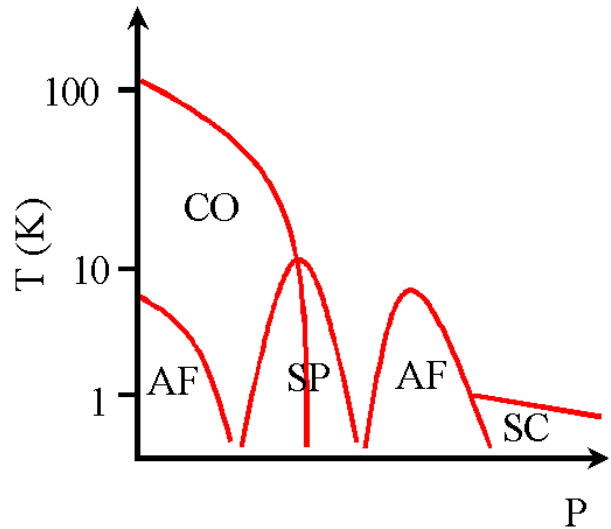}
\caption{\label{TMTTFpd} A proposed generic phase diagram for the TMTTF and TMTSF salts. }
\end{figure}

To conclude, we offer a generic phase diagram for the TMTTF and TMTSF salts that incorporates our observations for the 
$SbF_6$ salt. At low chemical pressure, the donor stacks are strongly dimerized, $T_{CO}$ is of the order $200K$, and the 
ground state is antiferromagnetic. Increasing pressure leads to a frustration of the CO resulting from a modified coupling 
to the counterion, and once it is sufficiently suppressed the ground state is singlet (Spin-Peierls) rather than 
antiferromagnetic. The observation that for \ssbf\, $dT_N/dP<0$, is presumably associated with a competition with this 
non-magnetic ground state found at higher pressure. Further pressurization leads to the familiar sequence: another AF 
state and superconductivity. It is reasonable to ask whether this second, higher-pressure, AF state is a re-entrance of 
the phase described in Fig. \ref{SsbfPD}, or whether it is a distinctly different symmetry breaking. The salts 
$(TMTTF)_2Br$ and $(TMTSF)_2PF_6$ should provide the answer, as they are understood to be representative of that portion 
of the phase diagram where the second AF phase appears. X-ray scattering results \cite{Pouget1997} from the two materials 
provide evidence for a coexistence of weak charge and bond modulations, suggesting an even richer evolution of the phases 
than is presented in Fig. \ref{TMTTFpd}.

\begin{acknowledgments}
The research was supported by the National Science Foundation under grant number DMR-0203806. The authors are grateful for 
conversations with S. Brazovskii, S. Mazumda, P. Monceau, and H. Seo.
\end{acknowledgments}


\begin{thebibliography}{24}
\bibitem[{app()}]{approximation}
\bibinfo{note}{Two inequivalent, coupled $I=1/2$ spins produce a spectrum with
  four lines. The Hamiltonian is given by
  $H=\omega_1I_{1z}+\omega_2I_{2z}+\omega_{12}I_{1z}I_{2z}$, and the lines
  appear at frequences $\sqrt{\omega_1^2+\omega_{12}^2}\pm\omega_{12}/2$,
  $\sqrt{\omega_2^2+\omega_{12}^2}\pm\omega_{12}/2$.}
\end{thebibliography}

\end{document}